\documentclass{eptcs}
\providecommand{\event}{QAPL 2010} 
\providecommand{\volume}{??}
\providecommand{\anno}{2010}
\providecommand{\firstpage}{1}
\providecommand{\eid}{??}
\usepackage{breakurl}        

\usepackage[english]{babel}

\usepackage{alltt,amsmath,amssymb}
 \usepackage{stmaryrd}
\usepackage{array}
\usepackage{gastex}
\usepackage{ifthen}
 \usepackage{wrapfig}
 \usepackage{enumerate}
\usepackage{xspace}
\usepackage{galois}

\usepackage{aeguill}
\usepackage{dsfont}

\usepackage{pgf}
 \usepackage{tikz}
\usepackage{multirow}
\usepackage{subfigure}

\newtheorem{definition}{Definition}
\newtheorem{lemma}{Lemma}
\newtheorem{proposition}{Proposition}
\newtheorem{theorem}{Theorem}

\def\squareforqed{\hbox{\rlap{$\sqcap$}$\sqcup$}}
\def\qed{\ifmmode\squareforqed\else{\unskip\nobreak\hfil
\penalty50\hskip1em\null\nobreak\hfil\squareforqed
\parfillskip=0pt\finalhyphendemerits=0\endgraf}\fi}

\newenvironment{proof}{\paragraph{Proof.}}{\hfill \qed}

 \newcommand{\Natural}{\mathbb{N}}
 \newcommand{\Real}{{\mathbb{R}}}
 \newcommand{\Realc}{\overline{\Real}}
 \newcommand{\Rational}{{\mathbb{Q}}}

\newcommand{\evInt}[1]{\mathop{\mathit{Int}}_{#1}^{[2]}}

 \newcommand{\ammaxshort}{(\Realc,\tt{max},+)}
 \newcommand{\amminshort}{(\Realc,\tt{min},+)}

 \newcommand{\state}{\Sigma}
 \newcommand{\sconc}{\sigma}
 \newcommand{\sabs}{\sigma^\sharp}
 \newcommand{\uco}{\pi}

 \newcommand{\p}{p}

\newcommand{\quant}{Q}

 \newcommand{\semtr}[1]{ \llbracket #1 \rrbracket_{\mathit tr}}
 
 \newcommand{\tr}{\mathop{\mathit tr}}
 \newcommand{\gc}{\mathop{\mathit gc}}
 
 \newcommand{\abstM}{M^{\sharp}}
 \newcommand{\step}{\mathop{\rightarrow^.}}
\newcommand{\AbState}{\state^\sharp}
 \newcommand{\abstdomain}{\state^\sharp} 
 \newcommand{\concdomain}{\state}        
 \newcommand{\abstelt}{\sigma^\sharp} 
 \newcommand{\concelt}{\sigma}        
\newcommand{\atoms}{\mathop{\mathcal{A}}}

\newcommand{\llift}[1]{{#1}^\uparrow}
\newcommand{\rlift}[1]{\overline{#1}}

\newcommand{\alphaup}{\rlift{\alpha}}
\newcommand{\BA}{\mathop{\mathcal{B}}(A)}
\newcommand{\BAup}{\rlift{\BA}}
\newcommand{\bc}{\rlift{b}}
\newcommand{\ba}{\rlift{ba}}
 \newcommand{\alphalift}{\rlift{\alpha_1}}
 \newcommand{\gammalift}{\rlift{\gamma_1}}

 \newcommand{\plusq}{\oplus}
 \newcommand{\bigplusq}{\bigoplus}
 \newcommand{\foisq}{\otimes}
 \newcommand{\bigfoisq}{\bigotimes}
 \newcommand{\zeroq}{\bot}
 \newcommand{\unq}{e}

 \newcommand{\mi}[1]{\mbox{\textit{#1}}}

 \newcommand{\fmi}[1]{\mathop{\mi{#1}}}

 \newcommand{\Id}{\fmi{Id}}

 \newcommand{\qm}[1]{\tilde{q}(#1)}
 \newcommand{\lgth}[1]{\left|#1\right|}

\title{Injecting Abstract Interpretations \\into Linear Cost Models}

\author{David Cachera\institute{ENS Cachan (Bretagne)/IRISA\\ Campus de Beaulieu\\ F-35042 Rennes,
  France} \email{david.cachera@irisa.fr}  
\and Arnaud Jobin\institute{Université Rennes 1/IRISA}\email{arnaud.jobin@irisa.fr}}

\def\titlerunning{Injecting Abstract Interpretations into Linear Cost Models}

\def\authorrunning{D.~Cachera \& A.~Jobin}

\begin{document}

\maketitle

\begin{abstract}
We present a semantics based framework for analysing the quantitative
behaviour of programs with regard to resource usage. We start from an
operational semantics equipped with costs. The dioid structure of the
set of costs allows for defining the quantitative semantics as a
linear operator. 
We then present an abstraction technique inspired from abstract
interpretation in order to effectively compute global cost information
from the program. Abstraction has to take two distinct notions of
order into account: the order on costs and the order on states.
We show that our abstraction technique provides a correct
approximation of the concrete cost computations.
\end{abstract}

\section{Introduction}

Static analyses are used to ensure qualitative properties on programs,
such as non-reachability of a given set of forbidden states. The
abstract interpretation theory encompasses many existing static
analyses and allows for systematically designing a variety of new ones
by defining abstract semantic domains and transfer functions adapted
to the problem under consideration. The main idea of abstract
interpretation is to replace concrete semantic computations (often
untractable or even uncomputable) by abstract ones which are
guaranteed to terminate, hopefully in reasonable time. 
  
In this paper, we are interested in analysing quantitative properties
of programs pertaining to the use of resources (time, memory, \ldots).
The computation of quantitative properties of program behaviours
suffers from the same drawbacks as their qualitative counterparts, and
thus needs adequate abstraction methods. The field of quantitative
software analysis has mainly concentrated on the analysis of
probabilistic properties, and the various corresponding models have
developed their own abstraction techniques. Modelling non functional,
but yet non probabilistic behaviour of programs has received less
attention.

Our starting point is an operational model of program execution where
the cost of each computational step is made explicit.  We take as
starting point a standard small-step operational semantics expressed
as a transition relation between states extended with \emph{costs}
associated to each transition. The set of costs is given a dioid (or
idempotent semiring) structure with two operators: a ``product''
operator that combines the costs along an execution path, and a
``sum'' operator that combines costs coming from different paths. This
allows for recasting the operational semantics into a framework of
linear operators over a vectorial structure, namely the moduloid of
vectors of costs indexed over the set of states.

Seeing the semantics of a program as a linear operator allows to
benefit from the nice algebraic properties of these operators. In particular, we
are able to easily define two notions of cost for a whole program
execution: a global cost from input to final states, meaningful only
if the program terminates, and a more interesting notion of long-run
cost, that corresponds to the maximum average of costs accumulated
along a cycle of the program semantics and provides an
over-approximation of the average cost per transition of long
traces. This latter notion is particularly interesting for the
analysis of programs with cyclic behaviour (such as reactive systems)
in which the asymptotic average cost along cycles, rather than the
global cost of the entire execution, is of interest.

Usual abstract interpretations are defined using Galois connections on
partially ordered structures, generally assuming the existence of a
complete lattice structure for concrete and abstract semantic
domains. In our model, we already have a notion of partial order, that
is the order on costs induced by the summation operator of the
dioid. This order is easily extended pointwise to vectors indexed over
states. If we do not assume any additional lattice order on states, we are able to
define a simple notion of partition based abstraction. This
abstraction technique has been developed
in~\cite{SotCacJen06,CacJenJobSot08}, and is suitable for
simple analyses that consist in ``forgetting'' information when going
from concrete to abstract states. If we want to use more elaborate
abstractions, and in particular reuse the classical abstractions of
standard abstract interpretation theory, we have to find an
abstraction technique that copes with two distinct notions of order:
the dioid order on costs, and the lattice order on states.  The
present paper addresses more specifically that question.

This paper is structured as follows.  Section~\ref{sec:linearsem}
defines the quantitative operational semantics of a program as a
linear operator, and gives the precise definition of cost dioid.
Section~\ref{sec:glrc} defines the notions of global and long-run cost
that can be extracted from the operational semantics.
Section~\ref{sec:gcres} gives the general definition of Galois
connection that is used in abstract interpretation theory, and shows
its relation with the notion of residuation that is used in our dioid
context.  Section~\ref{sec:pba} recalls the main results of partition
based abstractions, and shows the limitations of this technique.
Section~\ref{sec:lai} shows how abstractions can de designed that
respect both the dioid order of costs and the lattice structure of
states.  Section~\ref{sec:conclusion} gives related work and
concluding remarks.

\section{Linear operator semantics}
\label{sec:linearsem}

Transitions of the semantics are supplied with {\em quantities\/} (or
{\em costs\/}) depending on the accessed states. We consider as
semantic model a countable set of states $\state$, and define a program as a
transition system $P=\langle\state,\rightarrow^.,I,F\rangle$, where
$I$ is a set of initial states and $F$ a set of final states, without
referring to any particular syntax. The quantitative operational
semantics of $P$ is defined by the transition relation $\step
\subseteq \state \times \state \rightarrow \quant$ where a transition
\( \sigma \rightarrow^{q} \sigma' \) denotes a transition from state
$\sigma$ to state $\sigma'$ at cost $q$. The cost $q$ is function of
$\sigma$ and $\sigma'$, and the structure of the set $\quant$ of costs
will be made precise in the next subsection.

The trace semantics of $P$ is defined as follows.
\[\semtr{P} = \{ 
\sigma_0 \rightarrow^{q_0} \ldots \sigma_{n-1} \rightarrow^{q_{n-1}}
\sigma_{n} \mid
\sigma_0 \in I,
\sigma_i \rightarrow^{q_i} \sigma_{i+1}\}
\]

In the remainder of this section, we explain in more details the
structure we chose for sets of costs, namely cost 
dioids, before showing how the quantitative operational semantics can
be seen as a linear operator over vectorial structures constructed
from cost dioids.

\subsection{Cost dioid}
\label{subsec:dioids}

A transition $\sigma \rightarrow^{q} \sigma'$ of the quantitative
operational semantics states that a direct (one-step) transition from
$\sigma$ to $\sigma'$ costs $q$. These unitary transitions can be
combined into big-step transitions, using two operators: $\foisq$ for
accumulating costs and $\plusq$ to get a ``maximum'' of different
costs.  Costs can be defined in more general ways (for instance, one
could use a more general algebra of costs as
in~\cite{Aspinall:06:ResourceLogic}) but the present definition of
costs dioids covers a number of different costs and has interesting
computational properties, since it can be used within a linear
operator semantic framework, as presented in the next subsection.

The operator $\foisq$ on $\quant$ defines the global cost of a
sequence of transitions, $\sigma \rightarrow^{q_1} \ldots
\rightarrow^{q_n} \sigma'$ simply as $q = q_1 \foisq \ldots \foisq
q_n$. This is written $\sigma \stackrel{\p}{\Rightarrow}^q \sigma'$
where $\p$ is a sequence of states that has $\sigma$ (resp.\/
$\sigma'$) as first (resp.\/ last) state.

There may be several ways to reach a state $\sigma'$ from a state
$\sigma$, due to the presence of loops and non-determinism in the
semantics.  Let the set of possible paths be $\Pi_{\sigma,\sigma'} =
\{ \p\ \mid\ \sigma \stackrel{\p}{\Rightarrow}^{q_{\p}}
\sigma'\}$. The global cost between $\sigma$ and $\sigma'$ will be
defined, using the operator $\plusq$ on $\quant$, to be $q =
\bigplusq_{\p \in \Pi_{\sigma,\sigma'}} q_{\p}$. Formally, the two
operators have to fulfill the conditions of a (commutative) dioid.
\begin{definition}
\label{def:comdioid}
A {\em commutative dioid} is a structure $(\quant,\plusq,\foisq)$ such
that
\begin{enumerate}
\item Operator $\foisq$ is associative, commutative and has a neutral
  element $\unq$. Quantity $\unq$ represents a transition that costs
  nothing.
\item Operator $\plusq$ is associative, commutative and has $\zeroq$
  as neutral element.  Quantity $\zeroq$ represents the impossibility
  of a transition.
\item $\foisq$ is distributive over $\plusq$, and $\zeroq$ is
  absorbing element for $\foisq$ $(\forall
  x.x\foisq\zeroq=\zeroq\foisq x=\zeroq)$.
\item The preorder defined by $\plusq$ ($a\leq b\Leftrightarrow\exists
  c : a\plusq c=b$) is an order relation (\/{\em i.e.\/}~it satisfies
  $a\leq b$ and $b\leq a\Rightarrow a=b$).
\end{enumerate}
\end{definition}
A classical result of dioid theory~\cite{GonMinEng}.  states that
$\plusq$ and $\foisq$ preserve the order $\leq$, i.e., for all $a,b,c
\in \quant$ with $a\leq b$, $a\foisq c\leq b\foisq c$ and $a\plusq
c\leq b\plusq c$.

By nature, a dioid cannot be a ring, since there is an inherent
contradiction between the fact that $\plusq$ induces an order relation
and the fact that every element has an inverse for $\plusq$.

If several paths go from some state $\sigma$ to a state $\sigma'$ at
the same cost $q$, we will require that the global cost is also $q$,
{\em i.e.\/} we work with \emph{idempotent dioids}: $q \plusq q = q$ for all
$q$ in $\quant$.  Note that in an idempotent dioid $a\leq
b\Leftrightarrow a \plusq b=b$.  

The fact that sets of states may be
infinite, together with the use of residuation theory in
Section~\ref{sec:gcres} impose our structure to contain the
addition of any set of costs \footnote{This way, we define a complete
sup-semilattice over $\quant$.}.
\begin{definition}
An idempotent dioid is {\em complete\/} if it is closed with respect
to infinite sums, and the distributivity law holds also for an
infinite number of summands: for any set $X \subseteq \quant$, the
infinite sum $\bigoplus_{x \in X} x$ exists in the dioid and for all
$a \in \quant$, $ a \foisq (\bigoplus_{x \in X} x) = \bigoplus_{x \in
  X} (a \foisq x).$
\end{definition}

A complete dioid is naturally equipped with a top element, that we
shall write $\top$, which is the sum of all its elements.  We recall that
a complete dioid is always a complete lattice, thus equipped with a
meet operator $\wedge$~\cite{BacCohOlsQua92}.  The notion of long-run
cost we will define in Section~\ref{sec:glrc} relies on the
computation of an average cost along the transitions of a cycle. This
requires the existence of a $n$th root function.
\begin{definition}
A dioid $(\quant,\plusq,\foisq)$ is equipped with a {\em $n$th root
  function} if for all $q$ in $\quant$, equation $X^n=q$ has a {\em
  unique\/} solution in $\quant$, denoted by $\sqrt[n]{q}$.
\end{definition}
A sequence containing $n$ transitions, each costing, on average,
$\sqrt[n]{q}$, will thus cost $q$. Some examples of $n$th root can be
found in Table~\ref{tab:samples}.  To be able to easily deal with the
$n$th root, we make the assumption that the $n$th power is
\emph{$\plusq$-lower-semicontinuous} ($\plusq$-lsc for short): for all 
$X \subseteq \quant$, $(\bigplusq_{x \in X} x)^n = \bigplusq_{x \in X}
x^n$.  This assumption and its consequences will be very useful for
the theorems relating long-run cost and trace semantics in
Section~\ref{sec:glrc}. Note that this equality remains true for
finite $X$ (in that case the $n$th power is said a {\em
  $\plusq$-morphism\/}).

The following definition summarizes the required conditions for our
structure.
\begin{definition}[Cost dioid]
A {\em\ cost dioid\/} is a {\em complete\/} and {\em idempotent\/}
commutative dioid, equipped with an $n$th root operation, where
 the $n$th power is $\plusq$-lsc.
\end{definition}

Although the definition of cost dioids may seem rather restrictive, we
have shown in~\cite{CacJenJobSot08} that many classes of dioids
found in the literature are indeed cost dioids. The table displayed on
Table~\ref{tab:samples} gives a non exhaustive example list of cost
dioids. The taxonomy is borrowed from~\cite{BacCohOlsQua92}: a dioid
is {\em selective\/} \footnote{The order induced by a selective dioid is total.} if for all $a,b$, $a \plusq b $ is either $a$ or
$b$, {\em double-idempotent\/} if both $\plusq$ and $\foisq$ are
idempotent, and {\em cancellative\/} if for all $a,b,c$, $a \foisq b =
a \foisq c$ and $a \neq \zeroq$ implies $b = c$.

\begin{table}
\[
\begin{array}{|l|l|c|c|c|}
\hline
 & \mbox{carrier set} & \plusq & \foisq & \sqrt[n]{q}\\
\hline
 & \Rational \cup \{+\infty,-\infty\} & \min & \max & q\\
\mbox{Double-} & \Real \cup \{+\infty,-\infty\} & \max & \min & q\\
\mbox{idempotent} & \mathcal{P}(S) & \cap & \cup & q\\
 & \mathcal{P}(S) & \cup & \cap & q\\
\hline
\mbox{Cancellative} & \Real_{+}^m \cup \{+\infty\} & \min & + & \frac{q}{n}\\
\hline
& \Real_{+} \cup \{+\infty\} & \max & \times & q^{\frac{1}{n}}\\
\mbox{Selective} & \Rational \cup \{+\infty,-\infty\} & \max & + & \frac{q}{n}\\
& \Real \cup \{+\infty,-\infty\} & \min & + & \frac{q}{n}\\
\hline
\end{array}
\]
\caption{Some examples of cost dioids}\label{tab:samples}
\end{table}

The most common examples of cost dioids are $\ammaxshort$ and
$\amminshort$,  where $\Realc$ stands for $\Real\cup\{-\infty,+\infty\}$.  The
  induced orders are, respectively, the orders $\leq$ and $\geq$ over
  real numbers, extended to $\Realc$ in the usual way. These dioids
  are at the basis of discrete event systems theory, from which we
  borrow the notion of long-run cost in Section~\ref{sec:glrc}.

\subsection{Semantics as linear operators over dioids}
\label{subsec:linop}

Thanks to the multiplication and addition operators of the cost dioid,
the set of one-step transitions can be equivalently
represented by a {\em transition matrix\/} $M\in
\mathcal{M}_{\state\times \state}(\quant)$ with
\[\begin{array}{l}
M_{\sigma,\sigma'}=\left\{\begin{array}{l}
q\mbox{ if }\sigma \rightarrow^q \sigma'\\
\zeroq\mbox{ otherwise}\end{array}\right.
\end{array}\]
Here, $\mathcal{M}_{\state\times \state}(\quant)$ stands for the set of
matrices with rows and columns indexed over $\state$, and values in
$\quant$. In the following, a program
$P=\langle\state,\rightarrow^.,I,F\rangle$ will be equivalently  denoted as 
$P=\langle\state,M,I,F\rangle$ where $M$ is the matrix associated to
$\step$.
 
The set $\mathcal{M}_{\state\times \state}(\quant)$ is naturally equipped with two operators
$\plusq$ and $\foisq$ in the classical way: operator $\plusq$ is
extended pointwise, and operator $\foisq$ corresponds to the matrix
product (note that the iterate $M^n$ embed the costs for paths of
length $n$). Recall that the dioid is complete, ensuring existence
of the sum for each coefficient of the product matrix.  The resulting
structure is also an idempotent and complete dioid. The order induced
by $\plusq$ corresponds to the pointwise extension of the order over
$\quant$: $M \leq M' \Leftrightarrow \forall i,j.M_{i,j} \leq M'_{i,j}$.  A
transition matrix may also be seen as a linear operator on the
moduloid $\quant(\state)$, which is the analogue of a vector space
using a dioid instead of a field for external multiplication.

 If $E$ is an idempotent dioid, then for any moduloid $V$ over $E$ the
addition operator $\plusq$ defined pointwise is also idempotent, and
thus defines a canonical order on V. As for vector spaces, if $n$ is a given integer,
$E^n$, set of vectors with $n$ components in $E$, is a moduloid. More
generally, a vector $u \in E(\Sigma)$, with 
$\lgth{\Sigma} =n$ can be seen as a function $\delta_u : [1,n]
\rightarrow E$. Since $Q$ is complete, we can generalize to the
infinite countable case: $\delta_u$ becomes a mapping from
$\Natural$ to $E$. The
matrix-vector product is defined by: $(M u)_i =
\bigplusq_{j=1}^{+\infty} \delta_M(i,j) \foisq \delta_u(j)$.  In this
paper, we will keep the matrix notation for the sake of simplicity,
even for an infinite set of indices.\vspace{-.3cm}

\section{Global and long-run cost}
\label{sec:glrc}
\subsection{Global cost}
\label{subsec:gc}

Let $M$ be the matrix representing the quantitative transitions of a
program $P$. Recall that $M^k$ summarizes the transition costs of all
paths of length $k$. The global cost is then defined by computing the
successive iterates of the transition cost matrix until a fixpoint is
reached.  The transitive closure $M^+$ thus contains all the
transitions costs from any state to any state.
\[M^+ = \bigplusq_{i=1}^{\infty} M^i\]
The global cost of a program is obtained by extracting the
input-output cost from this transitive closure.
\begin{definition}
The global cost of a program $P=\langle\state,\rightarrow^.,I,F\rangle$ is defined as
\[\gc(P) = \bigplusq \{M^{+}_{i,f} | i \in I, f \in F \}\]
\end{definition}

Recall that, since we work in a complete semiring, this transitive
closure is always defined.
The global cost is  related to the standard trace semantics by the
following result~\cite{SotCacJen06}.

\begin{theorem}
\begin{equation}
\label{trEquiv}\gc(P) = \bigplusq \{ \bigfoisq^{f-1}_{j=1} q_j 
\mid \sigma_1 \rightarrow^{q_1} \ldots \rightarrow^{q_{f-1}} \sigma_f
\in \semtr{P}, \sigma_f \in F \} 
\end{equation}
\end{theorem}

Unfortunately, if the only information we get is that the global cost
is equal to the top element of the dioid, this definition is of little interest. This is the case in
particular when the semantics contains cycles of non-null cumulative
cost, which frequently arises when matrix $M$ is  an
abstraction of the semantics, as developed in
Section~\ref{sec:pba}. The notion of global cost thus correctly
deals with terminating programs over a finite state space, but is
inappropriate for reactive systems. For this reason, we rather use
the notion of long-run cost.

\subsection{Long-run cost}
\label{subsec:lrc}

Intuitively, the {\em long-run cost\/} of a program represents a
maximal average cost over cycles of transitions.
The average cost of a finite path is defined as the arithmetical mean
(w.r.t.\ the $\foisq$ operator) of the costs labelling its
transitions. In other words, it is the $n$th root of the global cost
of the path, where $n$ is its length.  We write
$\qm{\p}=\sqrt[\lgth{\p}]{q(\p)}$ for the average cost of path
$\p$, where $q(\p)$ is the global cost of $\p$, and $\lgth{\p}$
its length. The ``maximum'' average cost of all cycles in the
graph will be the quantity we are interested in: this quantity will be
called {\em long-run cost\/}.  The following example illustrates these
notions on a simple graph.
\vspace*{-2mm}
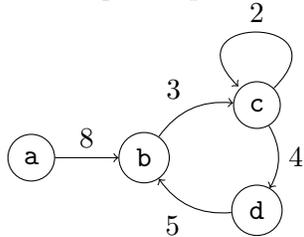
\begin{center} 
\begin{minipage}[h]{0.36\linewidth}
 \tikzstyle{state}=[circle,draw]
 \tikzstyle{pre}=[->]
 \tikzstyle{post}=[<-]
 \begin{tikzpicture}[auto,swap,node distance=15mm]
 \node[state] (a) {\tt a};
 \node[state] (b) [right of=a] {\tt b}
    edge[post] node {8} (a);
 \node[state] (c) [right of=b,yshift=7mm] {\tt c}
    edge[post,bend right] node {3} (b)
    edge[post,loop] node {2} (c);
 \node[state] (d) [right of=b,yshift=-7mm] {\tt d}
    edge[post,bend right] node {4} (c)
    edge[pre,swap,bend left] node {5} (b);
 \end{tikzpicture}
 \end{minipage}
\begin{minipage}[h]{0.53\linewidth}
Average cost of path {\tt abc} $=(8+3)/2=5.5$

Cycle {\tt bcdb} average cost $=(3+4+5)/3=4$

Cycle {\tt bccdb} average cost $=14/4=3.5$

Cycle {\tt cc} average cost $=2/1=2$

Long-run cost $=4$
\end{minipage}
\end{center}
\vspace*{-2mm}
The diagonal of matrix $M^k$ contains the costs of all cycles of
length $k$. If we add up all the elements on this diagonal, we get the
trace of the matrix. This observation gives rise to the following
definition.
 
\newcommand{\defLongrun}{
Let $P=\langle\state,M,I,F\rangle$ a program.
Let $R$ be $M$ restricted to the set of states, $\state_I$, reachable from $I$.
The long-run cost of program $P$ is defined as 
\[\rho(P) = \bigplusq_{k=1}^{\lgth{\Sigma_I}} \sqrt[k]{\tr R^k}
\quad \mbox{ where } \quad \tr R = \bigplusq_{i=1}^{\lgth{\Sigma_I}}
R_{i,i}.\] }
\begin{definition}\label{def:longrun}\defLongrun\end{definition}
Note that this definition is valid even for an infinite number of states, since
we work with complete dioids.  As an example, if we work in the dioid
$(\Realc, {\tt max},+)$,
$\rho(P)$ may represent the maximal average of time spent per instruction, where the
average is computed on any cycle by dividing the total time spent in the cycle
by the number of instructions in this cycle.  In the case of a finite set of
states, the long-run cost is computable, and we note in  passing that its
definition coincides with the definition of the maximum of eigenvalues of the
matrix, in the case of an irreducible matrix in an idempotent
semiring~\cite{BacCohOlsQua92}.

The following proposition~\cite{RR6338-v2} establishes in a more formal manner the link
between this definition of long-run cost and the cycles of the
semantics.  

\begin{proposition}\label{prop:cycles}
Let $\Gamma$ be the set of cycles
in  $\step$. Then $\rho(P) = \bigplusq_{c \in \Gamma} \qm{c}$.
\end{proposition}

As we aim at giving a characterisation of the asymptotic behaviour of
a program, we could have defined the long-run cost as the limit of the
average costs of all traces, instead of referring to cycles. The
drawback of this approach would be that this definition is not suitable for
computation, even if the set of states is finite. It is shown however
in~\cite{RR6338-v2} that those two notions coincide in a
restricted class of cost dioids and when the set of states is finite.

\section{Galois connections and residuation}
\label{sec:gcres}
The transition matrix representing a program is in general of infinite
dimension, so neither transitive closure nor traces can be computed in
finite time. Even if we deal with finitely machine-represented states,
the state space is in general too large for ensuring tractable
computations.  To overcome this problem, we define an abstract matrix
that can be used to approximate the computations of the original
matrix.  To prove the correctness of this approximation, we re-state
the classical abstract interpretation theory~\cite{Cousot} in terms of
linear operators over moduloids.  We first briefly recall a definition
of Galois connection that is  used in abstract interpretation.

\begin{definition}Let $(C,\leq_C)$ and $(D,\leq_D)$ be two partially ordered
  sets. Two mappings $\alpha: C \mapsto D$ (called abstraction function) and
  $\gamma: D \mapsto C$ (called concretization function) form a Galois
  connection $(C,\alpha,\gamma,D)$ iff:
\begin{itemize}
\item $ \forall c \in C, \forall d \in D, c \leq_C \gamma(d) \iff \alpha(c) \leq_D d,$ or equivalently
\item $\alpha$ and $\gamma$ are monotonic and $\alpha \circ \gamma \leq
  \mathit{Id}_{D}$ and $\mathit{Id}_{C} \leq \gamma \circ \alpha$
\end{itemize}
\end{definition}

The classical use of Galois connections considers complete lattices, but
their general definition is given on partially ordered sets. A
question that naturally arises is that of the existence of an analogous
notion relative to vectorial structures.  In the case of vector spaces
over the field of reals (more precisely, reals between 0 and 1
denoting probabilities), Di Pierro and Wiklicky~\cite{PiWi1} provide an elegant
solution by using the notion of Moore-Penrose
pseudo-inverse for bounded linear operators over Hilbert spaces. In our setting, we do not
have a field structure, but still benefit from a partial order
relation between vectors, namely the order induced by the $\plusq$
operators over vectors in a moduloid.
From a general point of view, the $\alpha$ and $\gamma$ mappings from
a Galois connection form a {\em pair of residuated
  maps\/}~\cite{MelSchStr86}. Applied to our dioid setting,
residuation theory can be restated as follows~\cite{BacCohOlsQua92}.

\begin{proposition}\label{prop:residu}
 Let $E$ and $F$ be two sets equipped with a complete partial order, $f$ a monotone mapping from
 $E$ to $F$. We call subsolution of equation $f(x)=b$ an element $y$
 such that $f(y) \leq b$. The following properties are equivalent.
\begin{enumerate}
\item For all $b \in F$, there exists a greatest subsolution to the
  equation $f(x)=b$.
\item $f(\bot_E) = \bot_F$, and $f$ is $\plusq$-lsc.\label{prop:residu:f_lsc}
\item There exists a monotone mapping $f^\dagger : F \to E$ which is
  upper\footnote{Upper semi-continuity is the analog of lower
  semi-continuity for the $\wedge$ operator.}
  semi-continuous such that $f \circ f^\dagger \leq \Id_F$
 and $\Id_E \leq f^\dagger \circ f$.
\end{enumerate}
Consequently, $f^\dagger$ is unique. When $f$ satisfies these
properties, it is said to be residuated, and $f^\dagger$ is called its
residual.
\end{proposition}
In our framework, the complete orders are the moduloid orders defined
pointwise from the cost dioid order.
If no additional order on the {\em set of states\/} is assumed, there is a
straightforward way to define residuable pairs of abstraction and
concretization functions on moduloids. This method of abstraction has
been developed in~\cite{SotCacJen06,CacJenJobSot08}, and we recall it in the next section to
show its limitations. 
If we start from an already existing abstraction function using a
lattice structure on states, we have to cope with two distincts
orders: the lattice order on states, and the dioid order on costs. We
thus have to define residuated pairs that take both orders into
account. This will be developed in Section~\ref{sec:lai}. 

\section{Partition-based abstraction}
\label{sec:pba}

We will first consider the simple case
where the abstraction is a mapping from concrete to abstract
states. This comes down to partitioning the set of concrete states
where equivalence classes are defined by abstract states. In this
section, $\state$ will denote a set of {\em concrete\/} states
and $\AbState$ a set of {\em abstract\/} states, with {\em no
  assumption\/} on the structure of these sets. In particular, they are
not supposed to be ordered.  An abstraction function $\alpha$ is thus a
mapping from $\state$ to $ \AbState$. In contrast, we  consider a cost dioid
$\quant$ with its partial order relation.

\subsection{Linear operator for abstraction}
\label{subsec:linabst}

If we now want to see the abstraction function as a linear abstraction
operator between the moduloids constructed on $\quant$ with indexes in
$\state$ and $\AbState$, respectively, we define the {\em linear lift\/}~\cite{CacJenJobSot08}
  of $\alpha$ as $\llift{\alpha} \in
\mathcal{M}_{\AbState\times \state}(\quant)$
by setting\footnote{Recall
that $e$ denotes the neutral element for $\foisq$.}
\[\begin{array}{l}
\llift{\alpha}_{\abstelt,\concelt} = \left\{\begin{array}{l}
\unq\mbox{ if }\alpha(\concelt) = \abstelt\\
\zeroq\mbox{ otherwise}
\end{array}\right.
\end{array}\]
In order to alleviate notations, $\leq$ will stand for the pointwise order defined on
$\mathcal{M}_{\state\times \state}(\quant)$ or
$\mathcal{M}_{\AbState\times \AbState}(\quant)$.
The pointwise orders defined on moduloids constructed over a complete
dioid are also complete. Moreover, as the abstraction function is
linear, it trivially fulfills requirements~\ref{prop:residu:f_lsc} of
Proposition~\ref{prop:residu}, and we get the following result~\cite{SotCacJen06} by
taking $\llift{\gamma}=(\llift{\alpha})^\dagger$.

\begin{theorem}\label{th:resid}
Let $\state$ and $\AbState$ be the domains of concrete and abstract states,
$\alpha$ a mapping from $\state$ to $\AbState$, and $\llift{\alpha} \in
\mathcal{M}_{\AbState\times \state}(\quant)$ the linear mapping obtained by lifting
$\alpha$. There exists a unique monotonic $\llift{\gamma}$ such that
\[ \llift{\alpha} \circ \llift{\gamma} \leq {\Id}_{\AbState\times\AbState}
\quad\mbox{and}\quad {\Id}_{\state\times\state}
\leq \llift{\gamma} \circ \llift{\alpha} \]
where ${\Id}_{\state\times\state}$ (resp.\/ ${\Id}_{\AbState\times\AbState}$) denotes the identity matrix in
$\mathcal{M}_{\state\times\state}(\quant)$ (resp.\/ $\mathcal{M}_{\AbState\times\AbState}(\quant)$).
\end{theorem}

The very simple form of abstraction we deal with up to now gives rise
to a very simple expression for $\llift{\gamma}$. Indeed, the unique
$\llift{\gamma}$ matching the requirements of Theorem~\ref{th:resid}
is the transpose matrix of $\llift{\alpha}$.

\subsection{Induced abstract semantics}
\label{subsec:abstsem}

Given a program $P$ over $\concdomain$, we want to define an abstract
transition system over the abstract domain $\abstdomain$ that is
``compatible'' with $P$, both from the point of view of its traces and
from the costs it  leads to compute.  The following definition of a
correct abstraction ensures that both global and long-run costs of
$P$ are correctly over-approximated
during the abstraction process.

\begin{definition}[Correct abstraction]
\label{def:correct}
Let $P=\langle\state,M,I,F\rangle$ be a transition system where $M \in
\mathcal{M}_{\concdomain\times \concdomain}(Q)$ and
$P^\sharp=\langle\AbState, \abstM,I^\sharp,F^\sharp\rangle$ be a
transition system over the abstract domain, with $\abstM \in
\mathcal{M}_{\abstdomain\times \abstdomain}(Q)$.  Let $\alpha$ be a
mapping from $\concdomain$ to $\abstdomain$.  The triple
$(P,P^\sharp,\alpha)$ is a correct abstraction from $\concdomain$ to
$\abstdomain$ if the three conditions 
 (1) $\llift{\alpha} \circ M \leq M^{\sharp} \circ
\llift{\alpha}$, (2) 
 $\{\alpha(\concelt)\mid\concelt\in I\}\subseteq
I^\sharp$
and (3) $\{\alpha(\concelt)\mid\concelt\in F\}\subseteq
F^\sharp$ hold.
\end{definition}

The classical framework of abstract interpretation gives a way to
define a best correct abstraction for a given concrete semantic
operator. In the same way, given an abstraction $\alpha$ and a
concrete semantics linear operator, we can define an abstract
semantics operator that is correct by construction, as expressed by
the following proposition~\cite{CacJenJobSot08}.

\begin{proposition}
  Let $\alpha$ be an abstraction from $\concdomain$ to $\abstdomain$,
  and $P=\langle\state,M,I,F\rangle$  a transition system  over the concrete
  domain. We set
  $P^\sharp=\langle \AbState, \abstM,I^\sharp,F^\sharp\rangle$ with
\[ M^\sharp = \llift{\alpha} \circ M \circ \llift{\gamma}
\quad\mbox{and}\quad I^\sharp= \{\alpha(\concelt)\mid\concelt\in I\}
\quad\mbox{and}\quad F^\sharp= \{\alpha(\concelt)\mid\concelt\in F\}\]
Then $(P, P^\sharp, \alpha)$ is a correct abstraction from
$\concdomain$ to $\abstdomain$. Moreover, given $P$ and $\alpha$,
$P^\sharp$ provides the best possible abstraction in the sense that if
$P'= \langle \AbState, M',I',F'\rangle$ is another correct
abstraction, then $M^\sharp \leq M'$ and $I^\sharp\subseteq
I'$ and $F^\sharp\subseteq F'$.
\end{proposition}

\subsection{Correctness of cost computations}
\label{subsec:corr}

The question that naturally arises is to know how global and
long-run costs are transformed by
abstraction. Theorems~\ref{th:corrgc} and~\ref{th:corrlrc} below
state that a correct abstraction gives an over-approximation of the
concrete global cost~\cite{RR6338-v2} and concrete long-run
cost~\cite{CacJenJobSot08}, respectively.

\begin{theorem}\label{th:corrgc}
If $(P,P^\sharp,\alpha)$ is a correct abstraction, then $\gc(P)\leq_Q \gc(P^\sharp)$.
\end{theorem}

\begin{theorem}\label{th:corrlrc}
If $(P,P^\sharp,\alpha)$ is a correct abstraction, then $\rho(P)\leq_Q \rho(P^\sharp)$.
\end{theorem}

 The proofs of these theorems rely on the fact that the correctness is
preserved when the concrete and abstract matrices are iterated
 simultaneously \cite{RR6338-v2}.

\subsection{Limitations}
\label{subsec:limits}

Partition based abstraction is well adapted to simple cases where
abstraction consists in ``forgetting'' information when going from the
concrete state to the abstract one. Let us take an example to
illustrate this fact. The concrete operational semantics of an object
oriented bytecode language considers states as tuples
$(h,(m,\mathit{pc},l,s)::\mathit{sf})$, where $h$ is the heap of
objects, and $(m,\mathit{pc},l,s)::\mathit{sf}$ is a call stack consisting of
\emph{frames} of the form $(m,\mathit{pc},l,s)$ where each frame contains a
method name $m$ and a program point $\mathit{pc}$ within $m$, a set of local
variables $l$, and a local operand stack~$s$ (see
\cite{siveroni04operational} for details on such an example).
Depending on the property the analysis wants to establish, a first
abstraction could define an abstract state as a simpler tuple
$(h,m,\mathit{pc},l,s)$, making the analysis context-insensitive. 
If we want to go further, we might want to abstract the heap $h$, which
is usually a mapping from locations to objects, by an abstract heap
mapping any location to the class of the corresponding object in the
concrete heap. Both of these abstractions are easily expressed by
abstraction functions  partitioning the set of concrete states, and
thus fit well the framework described above. 

In contrast, if we now want to abstract the values of local variables
by intervals, as is common in static analysis, we face two
problems. The first one is similar to a ``state explosion'' problem,
and the second one is related to the translation of the lattice order
of intervals into the moduloid structure over abstract states. Let us
explain both concerns in more details. Let $n$ be a natural number. We
denote by $\evInt{n}$ the set of  intervals with even bounds over
$\{-n,\dots,n\}$. The interval abstraction function
$\alpha_{\evInt{n}} \ : \mathcal{P}(\{ -n,\dots, n \}) \rightarrow
\evInt{n}$ maps a set of natural numbers $\{ i_1, \dots, i_r \}$ to
the interval $[ m - (m \mod 2), M + (M \mod 2) ]$ where $m = \min_{k
  \in \{ 1, \dots, r \}} i_k$ and $M = \max_{k \in \{ 1, \dots, r \}}
i_k$. If we lift $\alpha_{\evInt{n}}$ into a linear map as above, we
get a linear mapping from a moduloid of dimension $2^{2n+1}$ to a
moduloid of dimension $\frac{n(n+1)+2}{2}$. The corresponding matrix
is thus of size $\frac{n(n+1)+2}{2} \times 2^{2n+1}$. One could argue
that the subsets of $\{-n,\dots,n\}$ could canonically be represented
by a moduloid of dimension $2n+1$, each element contributing for one
dimension, and thus reducing the matrix size. For instance, if we fix
$n=2$, $\{-2\}$ is represented by $(\unq, \bot, \bot, \bot, \bot)^T$,
$\{2\}$ by $(\bot, \bot, \bot, \bot, \unq)^T$, $\{-2, 2\}$ by $(\unq,
\bot, \bot, \bot, \unq)^T$ etc. Let us now examine the abstract domain
of even intervals. The set of even intervals over $\{-2,\dots,2\}$ is
lifted to a moduloid of dimension $7$. For instance, $[-2]$ is
represented as $(\bot, \unq, \bot, \bot, \bot, \bot, \bot)^T$, and
$[2]$ as $(\bot, \bot, \bot, \unq, \bot, \bot, \bot)^T$, if we order
the intervals by increasing size and increasing lower bound.  We thus
should set $\alpha_{\evInt{2}}((\unq, \bot, \bot, \bot,
\bot)^T)=(\bot, \unq, \bot, \bot, \bot, \bot, \bot)^T$, and
$\alpha_{\evInt{2}}((\bot, \bot, \bot, \bot,\unq)^T)=(\bot, \bot,
\bot, \unq, \bot, \bot, \bot)^T$. Then $\alpha_{\evInt{2}}((\unq,
\bot, \bot, \bot, \unq)^T)=(\bot, \bot, \bot, \bot, \bot, \bot,
\unq)^T$, that is distinct from $\alpha_{\evInt{2}}((\unq, \bot, \bot,
\bot, \bot)^T)\oplus\alpha_{\evInt{2}}((\bot, \bot, \bot, \bot,
\unq)^T)$ which equals $(\bot, \unq, \bot, \bot, \unq, \bot,
\bot)^T$. In conclusion, we are not able to define
$\alpha_{\evInt{n}}$ as a linear operator as expected.  The problem
here comes from the fact that the structure of the abstract moduloid
totally forgets about the lattice structure of intervals. Defining a
residuable abstraction operator that respects the lattice structure of
abstract states is the main contribution of this paper, and is
developed in the next section.

\section{Lifting Abstract Interpretations}
\label{sec:lai}
In Section~\ref{sec:pba}, we have presented a way to lift any
abstraction function $\alpha: \concdomain \rightarrow \abstdomain$
into a linear mapping $\llift{\alpha} \in \mathcal{M}_{\abstdomain
  \times \concdomain}(Q)$, where domains $\concdomain$ and
$\abstdomain$ are not supposed to have a particular structure. In
order to benefit from the already existing abstractions provided by
the classical abstract interpretation theory, we show how to translate
them into our model. As abstract interpretation relies on lattices and
Galois connections, we will investigate in Section~\ref{subsec:abstop}
how these structures compare and are transposed to moduloids and
linear operators. Then, in Section~\ref{subsec:corrAI}, we will
investigate a new notion of correctness for this construction.

\subsection{Abstraction operator}
\label{subsec:abstop}
So far, the way we lift an abstraction represents a state $\sigma$ of
$\concdomain$ by a vector $(\bot, \dots, \bot, \unq, \bot, \dots,
\bot)^T$ where $\unq$ appears in the $\sigma$-place (recall $\concdomain$
is countable). The set of concrete states $\Sigma$ is thus represented
using the moduloid $\llift{\Sigma} = (\{\bot, \unq\}^{|\Sigma|}, \plusq,
\foisq)$. Now, if we assume that $\concdomain$ is a lattice, this lifting
unfortunately forgets about the ordered structure of
$\concdomain$. This is regrettable because $\llift{\Sigma}$
naturally has a lattice structure given by the $\plusq$ operator. Thus, a
natural issue is to translate $\Sigma$ and $\Sigma^{\sharp}$ into
moduloids while preserving their respective lattice orders. This
property of morphism between orders will be
referred to as the lift-order property in the remainder of this section.


\subsubsection{Lifting a Galois connection into a linear mapping}

Abstract interpretation often considers Galois connections $B
\galois{\alpha}{\gamma} A$ where $B$ is a powerset\footnote{Powersets are
  naturally equipped with a particular structure of complete lattice
  called boolean lattice \cite{DavPri}.} representing the concrete
semantic domain, and $A$ is a complete lattice representing the abstract
domain. In order to lift $\alpha$ into a linear mapping, we will focus
on how to lift-order these particular structures. The easy case
naturally is the one of boolean lattices.

\paragraph{Lift-ordering  boolean lattices.}

A boolean lattice $B$ is generated by its set of atoms
$\atoms(B)$, corresponding to the singletons in the case of a
powerset. Indeed, for each $b \in B$, $b = \vee \{ a \in
\atoms(B) \mid a \leq b \}$~\cite{DavPri}. Let us
code atoms $a$ as vectors $\llift{a}$ in $\{\bot,
\unq\}^{|\atoms(B)|}$ as previously (we note $\rlift{a} = \llift{a}$). Then, coding the other elements
will follow from  the use of $\oplus$.
\[ \rlift{b}  =  \oplus \{\rlift{a}  \mid a \leq b \}\]
We denote by $\rlift{B}$ the complete moduloid constructed this way
from $B$, where the $\plusq$ operator of $\rlift{B}$ matches the
$\cup$ operator of $B$ by construction.

Now that we have expressed boolean lattices as moduloids, we are able
to easily lift-order the abstraction function of a Galois connection
$B_1 \galois{\alpha}{\gamma} B_2$, where $B_1$ and $B_2$ are boolean
lattices. By lift-ordering these lattices, we obtain two moduloids
$(\rlift{B_1}, \plusq_1, \foisq_1)$ and $(\rlift{B_2}, \plusq_2,
\foisq_2)$. Since $\cup_i$ and $\plusq_i$ coincide,
and as $\alpha$ is a union morphism, its linear translation
$\rlift{\alpha}$ is defined
by its values on the basis vectors of  $\rlift{B_1}$, {\em i.e.\/} the
vectors coding atoms of $B_1$.  
\[
\begin{array}{lcrclcr}
\alpha(\{b_1\} & \cup_1 & \{b_2\}) & = & \alpha(\{b_1\}) & \cup_2 & \alpha(\{b_2\})\\
& \updownarrow & & & & \updownarrow & \\
\rlift{\alpha}(\rlift{b_1} & \plusq_1 & \rlift{b_2}) & = & \rlift{\alpha}(\rlift{b_1}) & \plusq_2 & \rlift{\alpha}(\rlift{b_2})\\
\end{array}
\]

\paragraph{Lift-ordering complete lattices.}

In most of the cases, $A$ is not a powerset but a more general complete
lattice for which the vectorial translation is not so straightforward.  The
representation theorem of finite distributive lattices~\cite{DavPri}
asserts that any such lattice $A$  is isomorphic
to a lattice of sets. Thus, $A$ can be seen as a sublattice of a
given powerset, which we will denote by $\BA$. The previous coding applies to
$\BA$ and \textit{a fortiori\/} to $A$. However, the set of
vectors $\rlift{A}$ constructed this way no more has a structure of
complete moduloid, unlike $\BAup$.
This method provides a solution to the ``state explosion'' problem presented
in Section~\ref{sec:pba}. Nevertheless, our second problem remains
unsolved. Indeed, there is still no match between the $\plusq$ operator and
$\cup$, the join operator of the lattice. For instance, $[-2] \cup [2] =
[-2,2]$ and $\rlift{[-2]} \plusq \rlift{[2]} = (\unq, \zeroq, \unq)^T$ and
$\rlift{[-2,2]} = (\unq, \unq, \unq)^T$. This makes it impossible to
express $\rlift{\alpha}$ as a linear mapping, since for instance
$\rlift{\alpha} (\rlift{\{-2\}} \plusq \rlift{\{2\}}) = (\unq, \unq,
\unq)^T \neq \rlift{\alpha} (\rlift{\{-2\}}) \plusq \rlift{\alpha}
(\rlift{\{2\}}) = (\unq, \bot, \unq)^T$.
We thus have to weaken our requirement: in the following,
we choose to lift-order Galois connections into non linear, but still
residuable, mappings.

\subsubsection{Lifting a Galois connection into a residuable mapping}

Since $\BAup$ is a complete boolean lattice, we will decompose
$\rlift{\alpha}$ into a linear part from $\rlift{B}$ to $\BAup$, and a
projection from $\BAup$ into its sublattice $\rlift{A}$ we are
interested in, representing the vector encodings of elements of
$A$. Figure~\ref{schema} illustrates this decomposition.

\begin{figure}[ht]
\begin{center}
\subfigure[Example of a set lattice (even interval lattice on the set
  $\{-2, \dots , 2\}$) and its associated powerset (which is
  isomorphic to the powerset $\mathcal{P}(\{1, 2,3\}), \cup$)]{
  \scalebox{0.8}{
\begin{picture}(80,67)
  \gasset{Nframe=if,Nadjust=w,Nh=6,Nmr=0}
  \node[Nframe=y](bot)(40,5){$\emptyset$}
  \node[Nframe=y](a)(20,25){$[-2]$}
  \node[Nframe=y](b)(40,25){$[0]$}
  \node[Nframe=y](c)(60,25){$[2]$}
  \node[Nframe=y](d)(20,45){$[-2, 0]$}
  \node[dash={1.5}0, Nframe=y](e)(40,45){$\ \ \ \ $}
  \node[Nframe=y](f)(60,45){$[0, 2]$}
  \node[Nframe=y](top)(40,65){$[-2, 2]$}

  \drawbpedge[AHnb=0,ATnb=0](bot,0,0,a,0,0){}
  \drawbpedge[AHnb=0,ATnb=0](bot,0,0,b,0,0){}
  \drawbpedge[AHnb=0,ATnb=0](bot,0,0,c,0,0){}
  \drawbpedge[AHnb=0,ATnb=0](a,0,0,d,0,0){}
  \drawbpedge[dash={1.5}0, AHnb=0,ATnb=0](a,0,0,e,0,0){}
  \drawbpedge[AHnb=0,ATnb=0](b,0,0,d,0,0){}
  \drawbpedge[AHnb=0,ATnb=0](b,0,0,f,0,0){}
  \drawbpedge[dash={1.5}0, AHnb=0,ATnb=0](c,0,0,e,0,0){}
  \drawbpedge[AHnb=0,ATnb=0](c,0,0,f,0,0){}
  \drawbpedge[AHnb=0,ATnb=0](d,0,0,top,0,0){}
  \drawbpedge[dash={1.5}0, AHnb=0,ATnb=0](e,0,0,top,0,0){}
  \drawbpedge[AHnb=0,ATnb=0](f,0,0,top,0,0){}

  \gasset{ExtNL=y,NLdist=0,AHnb=0,ilength=-2}
  \nodelabel[NLangle= 190](bot){\scalebox{.8}{$\left(\begin{array}{c} \bot\\ \bot\\ \bot \\ \end{array}\right)$}}

  \nodelabel[NLangle= 190](a){\scalebox{.8}{$\left(\begin{array}{c} \unq\\ \bot\\  \bot \\ \end{array}\right)$}}
  \nodelabel[NLangle= 190](b){\scalebox{.8}{$\left(\begin{array}{c} \bot \\ \unq\\ \bot\\ \end{array}\right)$}}
  \nodelabel[NLangle= -10](c){\scalebox{.8}{$\left(\begin{array}{c} \bot \\ \bot \\ \unq\\ \end{array}\right)$}}

  \nodelabel[NLangle= 190](d){\scalebox{.8}{$\left(\begin{array}{c} \unq\\ \unq\\  \bot \\ \end{array}\right)$}}
  \nodelabel[NLangle= 190](e){\scalebox{.8}{$\left(\begin{array}{c} \unq \\ \bot\\ \unq\\ \end{array}\right)$}}
  \nodelabel[NLangle= -10](f){\scalebox{.8}{$\left(\begin{array}{c} \bot \\ \unq \\ \unq\\ \end{array}\right)$}}

  \nodelabel[NLangle= 175](top){\scalebox{.8}{$\left(\begin{array}{c} \unq \\ \unq \\ \unq\\ \end{array}\right)$}}

\end{picture}
}
\label{powerset}
}
\subfigure[Galois connection and its lift]{
\scalebox{0.8}{
\begin{picture}(80,30)
  \gasset{Nframe=if,Nadjust=w,Nh=6,Nmr=0}
  \node[Nframe=n](b)(10,25){$B$}
  \node[Nframe=n](a)(70,25){$A$}
  \node[Nframe=n](bf)(10,5){$\rlift{B}$}
  \node[Nframe=n](baf)(50,5){$\BAup$}
  \node[Nframe=n](af)(70,5){$\rlift{A}$}

  \drawedge[dash={1.5}0](b,bf){}
  \drawedge[dash={1.5}0](a,af){}
  \drawedge[syo=-1,eyo=-1,ELside=r](b,a){$\alpha$}
  \drawedge[syo=1,eyo=1,ELside=r](a,b){$\gamma$}
  \drawedge[syo=-1,eyo=-1,ELside=r](bf,baf){$\alphalift$}
  \drawedge[syo=1,eyo=1,ELside=r](baf,bf){$\gammalift$}
  \drawedge[syo=-1,eyo=-1,ELside=r](baf,af){$\uco$}
  \drawedge[syo=1,eyo=1,ELside=r](af,baf){$\iota$}
 
\end{picture}
}
\label{schema}
}
\subfigure[Abstraction matrix mapping subsets of $\{ -2, \dots, 2\}$ to even intervals]{
\begin{minipage}{0.8\textwidth}
$$
\left(
  \begin{array}{ccccc}
    \unq & \unq & \zeroq & \zeroq & \zeroq \\
    \zeroq & \unq & \unq & \unq & \zeroq \\
    \zeroq & \zeroq & \zeroq & \unq & \unq \\
  \end{array}
\right)
$$

\end{minipage}
\label{alphaint2}}
\caption{Lifting of Galois connections}
\end{center}

\end{figure}

The linear part of $\rlift{\alpha}$, denoted by $\alphalift$ is defined as in
the case of a connection between two boolean lattices: $\alphalift$ is
defined on the set of atoms of $B$ by $\alphalift(\rlift{b}) =
\rlift{{\alpha(b)}}$ where $b$ is an atom of $B$, and then extended
to $\rlift{B}$ by linearity. As an example, Figure~\ref{alphaint2}
shows the abstraction matrix for the abstraction by even intervals,
for $n=2$. Element $\{-1\}$ of the concrete domain is mapped to
interval $[-2,0]$ of the abstract domain. Thus, $\alphalift$ maps
atom $\rlift{\{-1\}}$ to $\rlift{[-2,0]}$ which is the sum of atoms
  $\rlift{[-2]}$ and $\rlift{[0]}$.

This linear mapping is then composed with a projection $\uco$ in order
to yield a vector in $\rlift{A}$ corresponding to an element of the
(non boolean) lattice $A$. As we want to keep the lift-order property,
for all $x \in \BA$, $\uco(x)$ is defined as the smallest element $z
\in \rlift{A}$ such that $z \geq x$\footnote{The completeness property
  of $A$ and the morphism between the orders on $A$ and on its lifted
  version $\rlift{A}$ ensure the existence of this element.}. Note
that $\uco$ defined this way is an upper closure operator in $\BA$. On
our even interval abstraction example,
$\alphalift(\rlift{\{-2,2\}})=(\unq,\zeroq,\unq)^T$ is projected to
the top element $(\unq,\unq,\unq)^T$ of the abstract vector lattice.

As $\alphalift$ is a linear mapping between two complete moduloids, by
Proposition~\ref{prop:residu}, $\alphalift$ has a residual mapping $\gammalift$, {\em i.e.\/} $\alphalift
\circ \gammalift \leq \Id_{\BAup}$ and $\gammalift \circ \alphalift \geq
\Id_{\rlift{B}}$. Passing from $\rlift{A}$ to $\BAup$ is simply
done by a canonical injection $\iota$.

We finally prove the following property, that allows for defining a
pseudo-invertible lift of our initial Galois connection (the proof is
given in Appendix).

\begin{proposition}
\label{prop:galois_lift}
Mappings $\uco \circ \alphalift$ and $\gammalift \circ \iota$ as
defined above are such that $\uco \circ \alphalift$ is residuated and
$\gammalift \circ \iota$ is its residual, and thus form a Galois
connection between moduloids  $\rlift{B}$ and
$\rlift{A}$ seen as lattices.
\end{proposition}

By this construction, we are able to translate a Galois connection
$B\galois{\alpha}{\gamma} A$ into a residuable mapping written as the
composition of a linear mapping $\alphalift$ and an upper closure
operator $\uco$. We now check that this new construction of
abstraction operators preserves the over-approximation of cost
computations.

\subsection{Correctness of cost computations}
\label{subsec:corrAI}
The way of lifting abstraction functions in
Section~\ref{subsec:linabst} produced linear mappings. On the other
hand, the lift version of Section~\ref{subsec:abstop} gives only
residuable mappings. Unfortunately, the correctness of cost
computations intimately depends on the matricial character of the
abstraction function and seems difficult to establish for general residuable
mappings. 
Nonetheless, we can establish the correctness of cost computations in a
weaker way, using only the linear part $\alphalift$ of the residuable
abstraction $\uco \circ \alphalift$. We thus slightly change the
definition of a correct abstraction into a notion of
{\em correct linear abstraction\/}.
\begin{definition}[Correct linear abstraction]
\label{def:correctlinear}
Let $\rlift{B}$ and $\BAup$ be two moduloids of respective bases $\bc$
and $\ba$. Let $\quant$ be a cost dioid. Let $P=\langle
B,M,I,F\rangle$ a transition system with $M \in \mathcal{M}_{\bc\,
  \bc}(\quant)$ and $P^\sharp=\langle\BA,
\abstM,I^\sharp,F^\sharp\rangle$ be a transition system over the
abstract domain, with $\abstM \in \mathcal{M}_{\ba\,
  \ba}(\quant)$. Let
$\alphalift$ be a linear mapping from $\rlift{B}$ to $\BAup$. The
triple $(P,P^\sharp,\alphalift)$ is a correct linear abstraction from
$B$ to $\BAup$ if the three conditions (1) $\alphalift \circ M \leq M^{\sharp} \circ
\alphalift$, (2)  $\{\alpha(\concelt)\mid\concelt\in I\}\subseteq
I^\sharp$ and (3) $\{\alpha(\concelt)\mid\concelt\in F\}\subseteq
F^\sharp$ hold.
\end{definition}
In contrast with
Definition~\ref{def:correct} where we considered an abstraction function
$\alpha$ and stated the correctness using its lifted version
$\alphaup$,  we directly consider here the
abstraction function as a linear mapping between moduloids\footnote{For
  instance, this can be achieved by applying techniques of
  Section~\ref{subsec:abstop} on a Galois connection $B
  \galois{\alpha}{\gamma} A$}. As a consequence, we will prove a
notion of correctness that is
independent of the way  domains are lifted. As far as the global cost is concerned, this makes no difference since Lemma~\ref{lem:iterate} remains true for this notion. However, the correctness proof is
more difficult to achieve for the long-run cost, and will require an additional hypothesis
on  the cost dioid, namely it being selective.

 As the notion of long-run cost can be stated
without considering initial and final states, in what follows we use the notation of a
correct linear abstraction $(M, \abstM, \alphalift)$ to refer to the
inequality stated in  item~(1) of
Definition~\ref{def:correctlinear}.

 Theorem~\ref{correction_gc}
below states that a correct linear abstraction gives an overapproximation of
the global cost, while Theorem~\ref{correction_lrc} states the same result for the long-run cost.
\begin{theorem}
\label{correction_gc}
If $(M, M^\sharp, \alphalift)$ is a correct linear
abstraction, then $gc(M) \leq gc(\abstM).$
\end{theorem}
\begin{theorem}
\label{correction_lrc}
Let $\quant$ be a {\em selective} cost dioid\footnote{Recall that a
  dioid is {\em selective\/} if for all $a,b$, $a \plusq b $ is either
  $a$ or $b$}.  If $(M, M^\sharp, \alphalift)$ is a correct linear
abstraction, then $\rho(M) \leq \rho(\abstM).$
\end{theorem}
As mentioned above, proof of Theorem~\ref{correction_gc} is a direct consequence of Lemma~\ref{lem:iterate}. On the contrary, proof of Theorem~\ref{correction_lrc} requires four lemmas, whose proofs
are given in Appendix. 
\begin{lemma}
\label{lem:dev}
Let $(M, M^\sharp, \alphalift)$ be a correct linear abstraction and
($\sabs$, $\sconc$) $\in \ba \times \bc$. Then, we have:
 \begin{equation}
    \label{correction}
    \bigoplus_{\{c \in \bc| \sabs \leq \alphalift(c)\}} M_{c \sconc} \leq \bigoplus_{\{a \in \ba| a \leq \alphalift(\sconc)\}}  M^\sharp_{\sabs a}.    
  \end{equation}
\end{lemma}
Lemma~\ref{lem:dev} is quite
straightforward. Its proof consists in developing each member of the
correct linear abstraction inequality.
Lemma~\ref{lem:iterate} is not specific to this section and was borrowed
from the proof of Theorem~\ref{th:corrlrc}. 
\begin{lemma}
\label{lem:iterate}
Let $(M, M^\sharp, \alphalift)$ be a correct linear abstraction. Then,
for all $k \geq 1$, $(M^k, (M^\sharp)^k, \alphalift)$ is a correct
linear abstraction.
\end{lemma}

Lemma~\ref{lem:path} is the core of
Theorem~\ref{correction_lrc}. It establishes that every cycle of length
$k$ of the concrete graph represented by $M$ has a corresponding
abstract path of the same length $k$ and of higher cost. As mentioned
above, we will assume that the cost dioid is selective.
\begin{lemma}
  \label{lem:path}
  Let us assume a selective cost dioid $\quant$.  Let $(M, M^\sharp,
  \alphalift)$ be a correct linear abstraction. Then, for all
  $\sconc$ and $k \geq 1$, such that $M^k_{\sconc \sconc} \neq \bot$
  and for all ${\sabs_i} \in \ba$ appearing in the vector
  decomposition of $\alphalift (\sconc)$ on the basis $\ba$, there exists
  $\sabs_j \leq \alphalift (\sconc)$ such that $M^{\sharp k}_{\sabs_i
    \sabs_j} \geq M^k_{\sconc \sconc}$.
\end{lemma}

Finally, Lemma~\ref{lem:cycle} states that for every cycle of length
$k$ of the concrete graph $M$ there exists a cycle of potentially
higher length $kr$ in the abstract graph $\abstM$ and of higher
average cost.
\begin{lemma}
  \label{lem:cycle}
  Let $\quant$ be a selective cost dioid.  Let $(M, M^\sharp,
  \alphalift)$ be a correct linear abstraction.  Let $\sconc \in \bc$
  and $k \geq 1$ such that $M^k_{\sconc \sconc} \neq \bot$. We note
  $\alphalift (\sconc) = \sabs_1 \oplus \dots \oplus \sabs_s$. Then,
  there exist $1 \leq j, r \leq s$ such that:
  $$\sqrt[k]{(M^k)_{\sconc \sconc}} \leq \sqrt[kr]{(M^{\sharp kr})_{\sabs_j \sabs_j}}.$$
\end{lemma}



\section{Discussion and related work}
\label{sec:conclusion}
We have defined a quantitative counterpart of abstract interpretation
starting from an operational semantics where transitions are labelled
with costs of computations. The dioid structure of the set of costs
allows for defining concrete and abstract semantics as linear
operators between moduloids. We have presented two abstraction
techniques for relating concrete and abstract semantics. The first one
defines an abstraction function as a linear operator. It is usable for
simple cases of abstractions, but suffers from a state explosion
problem and is not suited for reusing standard abstract domains
provided by the abstract interpretation literature. The second
technique decomposes abstraction into a linear operator and a
projection operator, and establishes a link between our framework and
standard Galois connections. We have shown that both techniques
provide an over-approximation of concrete cost  computations.

This article follows~\cite{SotCacJen06,CacJenJobSot08}, where the
first abstraction technique was presented. It broadens our view of
quantitative  static analysis by allowing a reuse of classical
abstract domains used in qualitative static analyses. 

The present work is inspired by the quantitative abstract
interpretation framework developed by Di Pierro and
Wiklicky~\cite{PiWi1}.  We have followed their approach in modeling
programs as linear operators over a vector space, with the notable
technical difference that their operators act over a semiring of
probabilities whereas ours work with idempotent dioids. In Di Pierro
and Wiklicky's work, the relation with abstract interpretation is
justified by the use of the pseudo-inverse of a linear operator,
similar to a Galois connection mechanism, enforcing the soundness of
abstractions.  Our approach can be seen as intermediate between their
and classical abstract interpretation: on one hand, we use residuation
theory in order to get a pseudo-inverse for linear abstraction
functions; on the other hand, we benefit from the partially ordered
structure of dioids to give guarantees of soundness under the
assumption $\alpha\circ M\leq_D M^{\sharp}\circ\alpha$, which is a
classical requirement in abstract interpretation.  Another approach
for probabilistic abstract interpretation has been followed by
Monniaux~\cite{Mon1} for the analysis of imperative programs
containing random operators, where the semantics of a program is seen
as a mapping between probability distributions.  Note however that
none of the dioid approach and the probabilistic semiring approach can
generalise the other one, since there is an inherent contradiction
between being a ring and a dioid. Reconciling both frameworks would
require the definition of a more complex mathematical structure
equipped with all operators, with the difficulty of keeping all the
nice properties of the initial models.

Several other works make use of idempotent semirings for describing
quantitative aspects of computations, namely under the form of
constraint semirings~\cite{Bis04}, particularly under the name
of {\em soft constraints\/}. These have been used in the field of
Quality of Service~\cite{dfmpt05,San08}, in particular with systems
modelled by graph rewriting mechanisms~\cite{ht05}. In all these
approaches, the $\plusq$ and $\foisq$ operators of the constraint
semiring are used for combining constraints. Among these works, two
similar approaches deserve a particular attention, since they deal
with abstraction mechanisms.  Aziz~\cite{azi06} makes use of
semirings in a mobile process calculus derived from the
$\pi$-calculus, in order to model the cost of communicating
actions. He also defines a static analysis framework, by abstracting
``concrete'' semirings into abstract semirings of reduced cardinality,
and defining abstract semiring operators accordingly. Bistarelli {\em et al.\/}~\cite{BisCodRos02} define an abstract
interpretation based framework for abstracting soft constraint
satisfaction problems (SCSPs). As in Aziz's approach, they get an
abstract SCSP by just changing the associated semiring, leaving
unchanged the remainder of the structure. Concrete and abstract
semirings are related by means of a Galois insertion, which provides
correctness results. A major difference between these approaches and
ours is that they abstract the semiring and leave the system itself
unchanged, while we abstract the structures of states and keep the
same dioid. 

This paper tackles the problem of the linear operator approach
for modelling quantitative semantics. Even if we managed to get
residuated pairs for translating Galois connection into a linear
model, the correctness of cost computations for a lifted Galois
connection is defined only for its linear part, thus forgetting about
the final projection. One could argue
that this correctness is not adequate, since it does not deal with the
final abstract semantics but with an intermediate one. Recall however
that we aim at computing an over-approximation of the \emph{concrete}
long-run cost. Thus, the fact that the
``exact'' abstract semantics is obtained by a subsequent projection
does not really matter here.

An interesting avenue for further work would be to relax the correctness
criterion so that the abstract estimate is ``close'' to (but not necessarily
greater than) the exact quantity. For certain quantitative measures, a notion of
``closeness'' might be of interest, as opposed to the qualitative case where
static analyses must err on the safe side.

\bibliographystyle{eptcs}
\bibliography{QSA}

\appendix

\section{Proof of Proposition~\ref{prop:galois_lift}}
\setcounter{proposition}{3}

\begin{proposition}
Mappings $\uco \circ \alphalift$ and $\gammalift \circ \iota$ as
defined above are such that $\uco \circ \alphalift$ is residuated and
$\gammalift \circ \iota$ is its residual, and thus form a Galois
connection between moduloids  $\rlift{B}$ and
$\rlift{A}$ seen as lattices.
\end{proposition}

\begin{proof}
We first note that $\uco \circ \alphalift$ and $\gammalift \circ \iota$ are
monotonic by composition of monotonic mappings.  We then show that
$(\gammalift \circ \iota) \circ (\uco \circ \alphalift) \geq
\Id_{\rlift{B}}$: for all $a \in \rlift{B}$, $\uco (\alphalift(a))
\geq \alphalift(a)$ because $\uco$ is extensive. As $\gammalift$ is
monotonic and is the residual of $\alphalift$, we have $\gammalift \circ
\uco (\alphalift(a)) \geq \gammalift (\alphalift(a)) \geq a$.  We finally
show that $(\uco \circ \alphalift) \circ (\gammalift \circ \iota) \leq
\Id_{\BAup}$: as $\rlift{B}\galois{\alphalift}{\gammalift} \BAup$ is a Galois connection,
$\alphalift \circ \gammalift \circ \iota (x) = \alphalift \circ \gammalift (x)
\leq x$ for all $x \in \rlift{A}$. By applying the monotonic
function $\uco$ to each member of this inequality, we get $\uco
(\alphalift \circ \gammalift(x)) \leq \uco(x)$. As $x \in \rlift{A}$,
$\uco(x) = x$, which allows us to conclude the proof.
\end{proof}

\section{Proof of Theorem~\ref{correction_lrc}}

\setcounter{lemma}{1}

\begin{lemma}
Let $(M, M^\sharp, \alphalift)$ be a correct linear abstraction and
($\sabs$, $\sconc$) $\in \ba \times \bc$. Then, we have:
 \begin{equation}
      \bigoplus_{\{c \in \bc| \sabs \leq \alphalift(c)\}} M_{c \sconc} \leq \bigoplus_{\{a \in \ba| a \leq \alphalift(\sconc)\}}  M^\sharp_{\sabs a}.    
  \end{equation}
\end{lemma}

\begin{proof}
  Let first consider the left member of the correct linear abstraction inequality.
    $$
    \begin{array}{lclr}
      (\alphalift \circ M)_{\sabs \sconc} & = & \bigoplus_{c \in \bc} \alphalift_{\sabs c} \foisq M_{c \sconc}  & \\
      & = & \bigoplus_{\{c | \sabs \leq \alphalift(c)\}} M_{c \sconc} & \textrm{by definition of } \alphalift\\
    \end{array}
    $$ We note in the passing that inequality $\sabs \leq \alphalift(c)$
    is equivalent to the fact that the element $\sabs \in \ba$
    appears in the vector decomposition of $\alphalift (c)$ over the
    basis $\ba$.

We conclude the proof by developing the right member
    of the inequality:
    $$
    \begin{array}{lcl}
      (M^\sharp \circ \alphalift)_{\sabs \sconc} & = & \bigoplus_{a \in \ba} M^\sharp_{\sabs a} \foisq \alphalift_{a \sconc} \\
      & = & \bigoplus_{\{a | a \leq \alphalift(\sconc)\}}  M^\sharp_{\sabs a}\\
    \end{array}
    $$
\end{proof}

\begin{lemma}
Let $(M, M^\sharp, \alphalift)$ be a correct linear abstraction. Then,
for all $k \geq 1$, $(M^k, (M^\sharp)^k, \alphalift)$ is a correct
linear abstraction.
\end{lemma}

\begin{proof}
We proceed by induction over $k$. The property holds at rank $1$ by
hypothesis. If the property holds at rank $n$, it is also established
at rank $n+1$ by applying property at rank $1$ and by preservation of
the order in a dioid.
\end{proof}

\begin{lemma}
 
  Let $\quant$ be a selective cost dioid.  Let $(M, M^\sharp,
  \alphalift)$ be a correct linear abstraction.  Then, for all
  $\sconc$ and $k \geq 1$, such that $M^k_{\sconc \sconc} \neq \bot$
  and for all ${\sabs_i} \in \ba$ appearing in the vector
  decomposition of $\alphalift (\sconc)$ on the basis $\ba$
  ($\Leftrightarrow \sabs_i \leq \alphalift (\sconc)$), there exists
  $\sabs_j \leq \alphalift (\sconc)$ such that $M^{\sharp k}_{\sabs_i
    \sabs_j} \geq M^k_{\sconc \sconc}$.
\end{lemma}

\begin{proof}
  Let $\sconc \in \bc$ such that $M^k_{\sconc \sconc} \neq \bot$. We
  note $\alphalift (\sconc) = \sabs_1 \oplus \dots \oplus \sabs_s$ the
  vector decomposition of $\alphalift (\sconc)$ on the basis
  $\ba$.\\ We apply inequality (\ref{correction}) to $(\sabs_1,
  \sconc)$, \dots, $(\sabs_s, \sconc)$, $M^k$ and $M^{\sharp k}$.\\ As
  for all $i$, $\sconc$ belongs to $\{c | \sabs_i \leq
  \alphalift(c)\}$, we get:
  $$
  \begin{array}{lclcl}
    M^k_{\sconc \sconc} & \leq & M^{\sharp k}_{\sabs_1 \sabs_1} \oplus 
\dots \oplus M^{\sharp k}_{\sabs_1 \sabs_s} & = & M^{\sharp k}_{\sabs_1 \sabs_{m_1}}\\
     & \vdots & & \vdots & \\
    M^k_{\sconc \sconc} & \leq & M^{\sharp k}_{\sabs_s \sabs_1} \oplus 
\dots \oplus M^{\sharp k}_{\sabs_s \sabs_s} & = & M^{\sharp k}_{\sabs_s \sabs_{m_s}}\\
  \end{array}
  $$ where $m_i$ denotes the index of the greatest element of the
  right member of the inequality (recall that we demand the dioid to
  be selective). Thus, for all $\sabs_i$, $M^{\sharp k}_{\sabs_i
    \sabs_{m_i}} \geq M^{k}_{\sconc \sconc}$.
\end{proof}

\begin{lemma}
   Let $\quant$ be a selective cost dioid.  Let $(M, M^\sharp,
  \alphalift)$ be a correct linear abstraction.  Let $\sconc \in \bc$
  and $k \geq 1$ such that $M^k_{\sconc \sconc} \neq \bot$. We note
  $\alphalift (\sconc) = \sabs_1 \oplus \dots \oplus \sabs_s$. Then,
  there exist $1 \leq j, r \leq s$ such that:
  $$\sqrt[k]{(M^k)_{\sconc \sconc}} \leq \sqrt[kr]{(M^{\sharp kr})_{\sabs_j \sabs_j}}.$$
\end{lemma}

\begin{proof}
Applying Lemma~\ref{lem:path}, there exist $(m_i)_{1 \leq i \leq s}$,
elements of $\llbracket 1, s \rrbracket$ such that, for all $i$,
$M^{\sharp k}_{\sabs_i \sabs_{m_i}} \geq M^k_{\sconc \sconc}$. It
implies that every edge ($\sabs_i$, $\sabs_{m_i}$) of the graph
$M^{\sharp k}$ has a non-zero cost ($\neq \bot$). Every vertice of the
graph $M^{\sharp k}$ restricted to the vertices $\{ \sabs_1, \dots ,
\sabs_s \}$ has at least one leaving edge. We deduce from this that there
exists a cycle in this restricted graph. Thus, there is $1 \leq j \leq
s$ such that $M^{\sharp k}_{\sabs_j \sabs_{l_1}} \geq M^k_{\sconc
  \sconc}$, \dots, $M^{\sharp k}_{\sabs_{l_r} \sabs_j} \geq
M^k_{\sconc \sconc}$ for an appropriate $r \in \llbracket 1, s
\rrbracket$. By order preservation,  we get:
$$ aux = M^{\sharp k}_{\sabs_j \sabs_{l_1}} \foisq \dots \foisq
M^{\sharp k}_{\sabs_{l_r} \sabs_{j}} \geq (M^k_{\sconc \sconc})^r.$$
By definition of the diagonal elements of $M^{\sharp kr}$, we get that
$M^{\sharp kr}_{\sabs_j \sabs_j} \geq aux \geq (M^k_{\sconc
  \sconc})^r$. We recall that the $kr$th power is a monotonic function. Thus, it suffices to apply it to each side of the inequality to get the wanted
result.
\end{proof}

Now, we can establish Theorem~\ref{correction_lrc}.
\begin{proof}
  By applying Lemma~\ref{lem:cycle}, we get that for each cycle $c$ of $M$ there exists a cycle $c^\sharp$ of $M^\sharp$ of higher average cost ($\tilde{q} (c^\sharp) \geq \tilde{q} (c)$). Thus,
$$ \rho(M) = \bigoplus_{c \textrm{ cycle of } M} \tilde{q}(c) \leq \bigoplus_{c^\sharp \textrm{ cycle of} M^\sharp} \tilde{q}(c^\sharp) = \rho (M^\sharp).$$
\end{proof}                                      

\end{document}